\documentclass[twoside,leqno,twocolumn]{article}

\usepackage[letterpaper]{geometry}

\usepackage{ltexpprt}

\usepackage{hyperref}
\usepackage{color}

\usepackage{algorithmicx} 
\usepackage{algpseudocode} 
\usepackage{graphicx}
\usepackage{fancyhdr}
\usepackage{amsmath}
\usepackage{booktabs}
\usepackage{cleveref}
\usepackage{makecell}
\usepackage{subfigure}

\title{Fast and Robust Hexahedral Mesh Optimization via Augmented Lagrangian, L-BFGS, and Line Search}

    \author{Hua Tong\thanks{Department of Mechanical Engineering, Carnegie Mellon University.}
    \and Yongjie Jessica Zhang\thanks{Department of Mechanical Engineering, Carnegie Mellon University.}}    

\date{\today}
\begin{document}
\maketitle


\fancyfoot[R]{\scriptsize{Copyright \textcopyright\ 2025 by SIAM\\
Unauthorized reproduction of this article is prohibited}}





\begin{abstract} \small\baselineskip=9pt
We present a new software package, ``HexOpt,'' for improving the quality of all-hexahedral (all-hex) meshes by maximizing the minimum mixed scaled Jacobian-Jacobian energy functional, and projecting the surface points of the all-hex meshes onto the input triangular mesh. The proposed HexOpt method takes as input a surface triangular mesh and a volumetric all-hex mesh. A constrained optimization problem is formulated to improve mesh quality using a novel function that combines Jacobian and scaled Jacobian metrics which are rectified and scaled to quadratic measures, while preserving the surface geometry. This optimization problem is solved using the augmented Lagrangian (AL) method, where the Lagrangian terms enforce the constraint that surface points must remain on the triangular mesh. Specifically, corner points stay exactly at the corner, edge points are confined to the edges, and face points are free to move across the surface. To take the advantage of the Quasi-Newton method while tackling the high-dimensional variable problem, the Limited-Broyden-Fletcher-Goldfarb-Shanno (L-BFGS) algorithm is employed. The step size for each iteration is determined by the Armijo line search. Coupled with smart Laplacian smoothing, HexOpt has demonstrated robustness and efficiency, successfully applying to 3D models and hex meshes generated by different methods without requiring any manual intervention or parameter adjustment.
\end{abstract}

\section{Introduction}
\label{sec:1}

Hexahedral (hex) mesh generation plays an important role in solving partial differential equations in multiple fields such as computer graphics, medical modeling, and engineering simulations \cite{zhang2016geometric}. Compared to tetrahedral meshes, hex meshes are generally preferred due to their higher accuracy, fewer element counts, and greater reliability \cite{benzley1995comparison, shepherd2006quality}. Despite these recognized benefits, automatic generation of high-quality and conforming hex meshes remains a significant challenge \cite{owen1998survey, zhang2013challenges,zhang2010automatic}. The generation of high-quality hexahedral meshes typically involves (1) initial mesh generation with connectivity designed to fit the input geometry; (2) vertex position modification to optimize the mesh quality and geometry fitting \cite{owen1998survey}. The initial meshes that serve as input to step 2 often contain poorly shaped and even inverted elements. On one hand, eventually all the inverted elements need to be eliminated, because even a single inverted, or non-convex element, makes a mesh unusable for simulation. On the other hand, the rigid structure of hex elements complicates local adjustment strategies, unlike the more flexible quadrilateral or tetrahedral meshes \cite{schneiders2000algorithms}. 

Due to the aforementioned reason, hex mesh optimization remains an active and challenging research area \cite{knupp2001hexahedral, shepherd2008hexahedral, pietroni2022hex}. It involves improving the quality of the worst elements and aligning the quadrilateral surface with the input triangular boundary. Many algorithms have emerged to improve mesh quality. One such method, Laplacian smoothing, is both straightforward and effective, repositioning vertices to the centroid of their adjacent vertices \cite{canann1998approach}. While cost-efficient and easy to implement, this technique risks inverting neighboring elements. To mitigate this problem, optimization-based strategies are proposed to evaluate and improve the quality of elements neighboring a node \cite{zhang2006adaptive, zhang2009surface,zhang20053d}. A hybrid approach combining Laplacian smoothing with optimization, can balance between efficiency and robustness \cite{freitag1997combining, canann1998approach}. For non-manifold hex meshes in micro-structured materials, a specialized method utilizing a vertex categorization system integrated with pillowing, geometric flow, and optimization is proposed, addressing previous research limitations \cite{qian2010quality}.

An untangling scheme performs single and fast local linear programming and traverses through each vertex until the quality cannot be improved any more \cite{freitag2000local}. Such method has local convergence proof, whereas it gets trapped in local minimum when the local solution space is empty, and better local solution can only be achieved by simultaneously moving multiple vertices. Therefore, an edge-cone rectification method that combines local quadratic programming with global reconciliation is employed to achieve good practical performance \cite{livesu2015practical}. A similar Newton-Raphson-based method that maximizes the average scaled Jacobian is proposed \cite{ruiz2015simultaneous}, whereas this method does not consider fitting the geometry surface.

Several other iterative techniques have been proposed to gradually shift vertices towards boundaries while avoiding local adjustments that might result in hex flipping \cite{marechal2009advances, lin2015quality, qian2012automatic}. These methods are straightforward but may sometimes struggle to maintain precise geometry. A global deformation method has been shown to exhibit robust alignment of the resultant mesh with the input surface through surface mapping. The fitting process is controlled using a Hausdorff distance threshold \cite{xu2018hexahedral}. However, this method sometimes fails to preserve exact geometry in their experimental results. A structure-aware geometric optimization method for hex meshes is proposed \cite{wang2021structure}, aiming to improve mesh quality by optimizing the positions of singular lines and parameterization in the base complex structure. However, it cannot optimize hex meshes with tangled meshes and struggles with meshes featuring high-valence edges. A three-stage optimization process for all-hex meshes is also proposed \cite{huang2022untangling}, which is effective in producing high-quality, inversion-free meshes, may incur additional computational time due to its sequential process.

In this paper, starting with a 3D closed manifold surface including any annotated sharp features, our method focuses on minimizing an objective function combining two widely recognized algebraic metrics, Jacobian and scaled Jacobian, which are rectified and scaled to quadratic measures. We conduct analysis of the scaled Jacobian function and effectively address its undesirable behavior in degenerate regions. To evaluate optimization performance, we compare two methods: the steepest descent with a fixed learning rate (adopted in the previous software package ``HybridOctree\_Hex'' \cite{tong2024hybridoctree_hex}) and L-BFGS with Armijo line search. For surface points, we compute their projection points on the triangle surface, which is the equality constraint. These equality constraints are handled using the AL method. Ultimately, we incorporate Laplacian smoothing to accelerate convergence and tackle situations where surface points get trapped in local minima. In the experiments, we eliminate self-intersections and obtain good minimum scaled Jacobians across all tested models, surpassing the current state-of-the-art results \cite{livesu2015practical}. Additionally, our innovative approach excels in accurately preserving intricate curved and sharp features.  To foster additional research and collaboration, we make available HexOpt source code, as well as a collection of generated meshes and their input-output data, accessible at \url{https://github.com/CMU-CBML/HexOpt}.

The paper is organized as follows: Section \ref{sec:2} presents the algebraic shape quality metric, focusing on the (scaled) Jacobian for hex elements. It also explores the detailed algorithms of the AL objective function, the L-BFGS, and the Armijo line search, as well as the pipeline pseudo-code. Section \ref{sec:3} exhibits meshing examples, demonstrating that our proposed method creates valid meshes composed of high-quality hexes. Lastly, Section \ref{sec:4} concludes the paper and provides insight into potential future research directions.

\section{Hex Mesh Optimization}
\label{sec:2}

\begin{figure*}
  \hspace{-16mm}
  \includegraphics[width=1.15\textwidth]{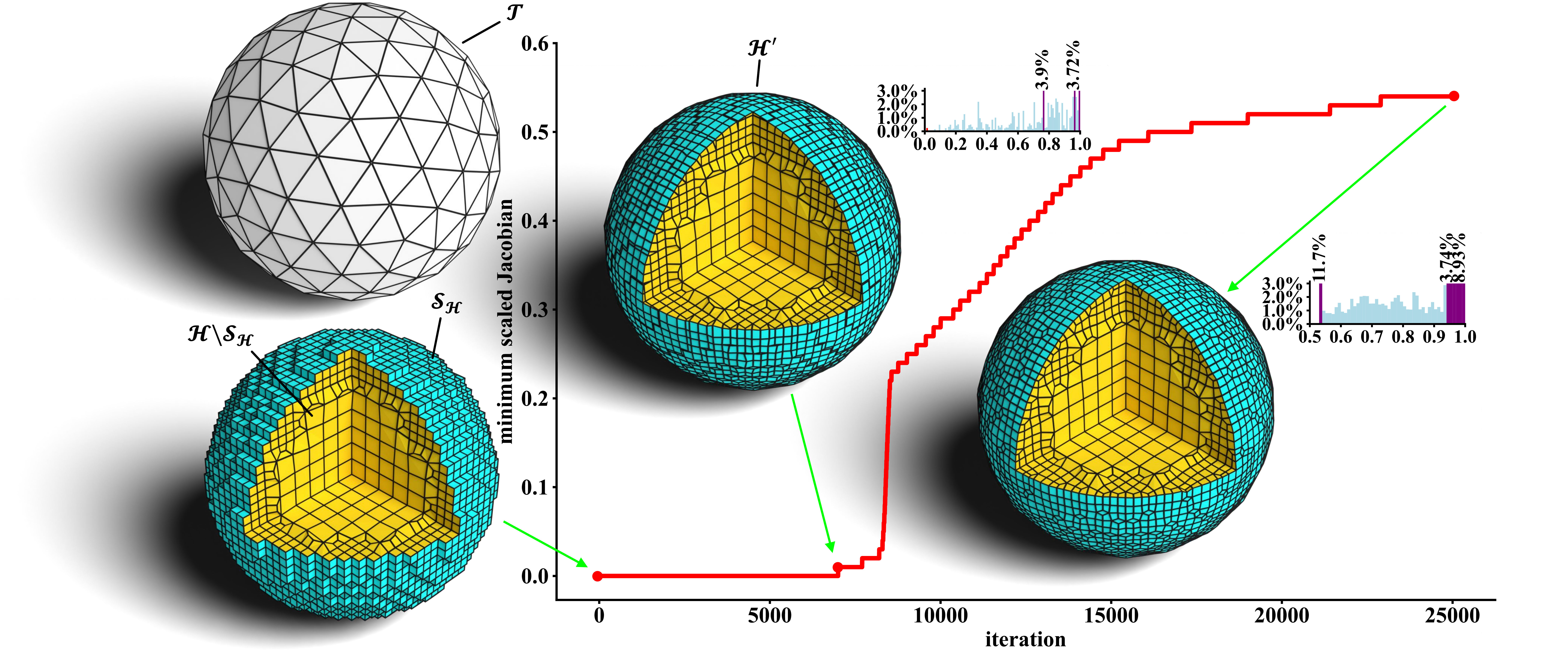}
  \vspace{-7mm}
  \caption{Hex mesh optimization process. The triangle surface $\mathcal{T}$ is shown in white. The quadrilateral surface $\mathcal{S}_\mathcal{H}$ is shown in blue. The hex core mesh $\mathcal{H}\setminus\mathcal{S}_\mathcal{H}$ is shown in yellow. The vertices are warped to minimize the objective function in Equation (\ref{equ:AL}). The middle stage mesh $\mathcal{H}'$ with $\mathcal{S}_{\mathcal{H'}}$ exactly fits to $\mathcal{T}$ with the minimum scaled Jacobian of 0.01. The minimum scaled Jacobian is increased by 0.01 every time with $\mathcal{S}_{\mathcal{H'}}$ exactly fitting to $\mathcal{T}$ until we cannot improve anymore, and we export the final mesh at the bottom right.}
  \label{fig:bigpicture}
\end{figure*}

The input is a watertight triangular mesh $\mathcal{T}$ annotated with sharp features and its corresponding hex mesh $\mathcal{H}$. The word ``watertight'' means that each edge is shared by exactly two faces. The surface of $\mathcal{H}$ is denoted as $\mathcal{S}_{\mathcal{H}}$. Although $\mathcal{S}_{\mathcal{H}}$ approximates the triangular mesh surface, it lacks accurate fitting. To address this issue, we adjust $\mathcal{H}$ to $\mathcal{H}'$ to ensure that $\mathcal{S}_{\mathcal{H}'}$ fits $\mathcal{T}$ while maintaining a high minimum scaled Jacobian. Figure \ref{fig:bigpicture} illustrates our mesh optimization process. The top left showcases the sphere-shaped surface geometry $\mathcal{T}$. The bottom left shows the initial core mesh $\mathcal{H}$ in yellow and its surface $\mathcal{S}_{\mathcal{H}}$ in blue. When the optimization begins, the gradient of the objective function measuring mesh quality and geometry fitting is calculated for each vertex in $\mathcal{H}'$, and the vertices are warped based on the approximated Hessian matrix. After some iterations, the vertices on $\mathcal{S}_{\mathcal{H}'}$ (the surface of $\mathcal{H}'$) first fit $\mathcal{T}$, and the minimum scaled Jacobian increases; see the middle picture. Subsequently, we continue to optimize both geometry fitting and mesh quality until we can no longer improve the minimum scaled Jacobian without deviating vertices on $\mathcal{S}_{\mathcal{H}'}$ from $\mathcal{T}$. The final optimized mesh is shown in the bottom right.

\subsection{Algebraic Quality Measures for Hex Elements}
\label{sec:2.1}

We adopt the scaled Jacobian to measure mesh quality \cite{knupp2006verdict}. Within each hex $h$, for every corner node $x$, three edge vectors are defined as $e_i=x_i-x$ $(i = 0, 1, 2)$. The Jacobian matrix at $x$ is constructed as $[e_0, e_1, e_2]$, and its Jacobian, $J(x)$, is the determinant of this matrix. We obtain the Scaled Jacobian, $SJ(x)$, when $e_0$, $e_1$, and $e_2$ are normalized to $\frac{e_i}{\lVert e_i \rVert_2}$. For the scaled Jacobian $SJ(h)$, we compute at the eight corners and the body center, and the hex scaled Jacobian is the minimum of these nine values. For the body center, $e_i$ ($i = 0, 1, 2$) is calculated using three vectors formed by pairs of opposite face centers. The scaled Jacobian value range is $[-1, 1]$.

\subsection{Mixed Scaled Jacobian and Jacobian}

Since we want to maximize the minimum scaled Jacobian of $\mathcal{H}$, the straightforward idea is to use the so-called Rectified Scaled Jacobian (\textit{ReSJ}) as the objective function:
\begin{equation}
\max\sum_{h\in\mathcal{H}}ReSJ(h, \Theta),
\label{equ:naive}
\end{equation}
where $\Theta > 0$ is the threshold for the minimum scaled Jacobian value,
\begin{equation}
ReSJ(h, \Theta) = \begin{cases}
SJ(h), & \text{if } SJ(h) \leq \Theta \\
\Theta. & \text{if } SJ(h) > \Theta
\end{cases}
\label{equ:ReSJ}
\end{equation}

\begin{figure*}
  \centering
  \subfigure[]{
    \includegraphics[width=0.308\textwidth]{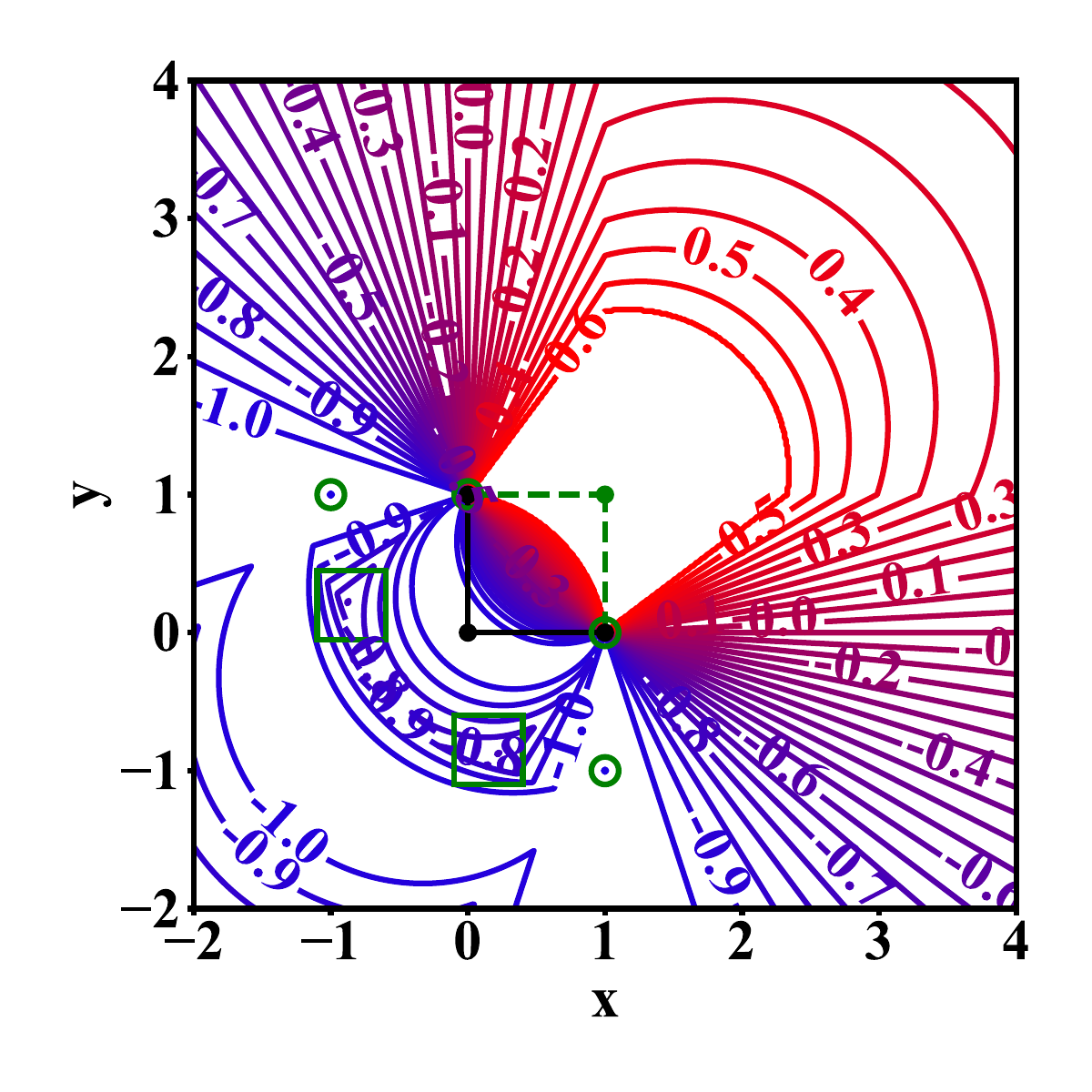}
    \label{fig:2DReSJ}  
  }
  \subfigure[]{
    \includegraphics[width=0.308\textwidth]{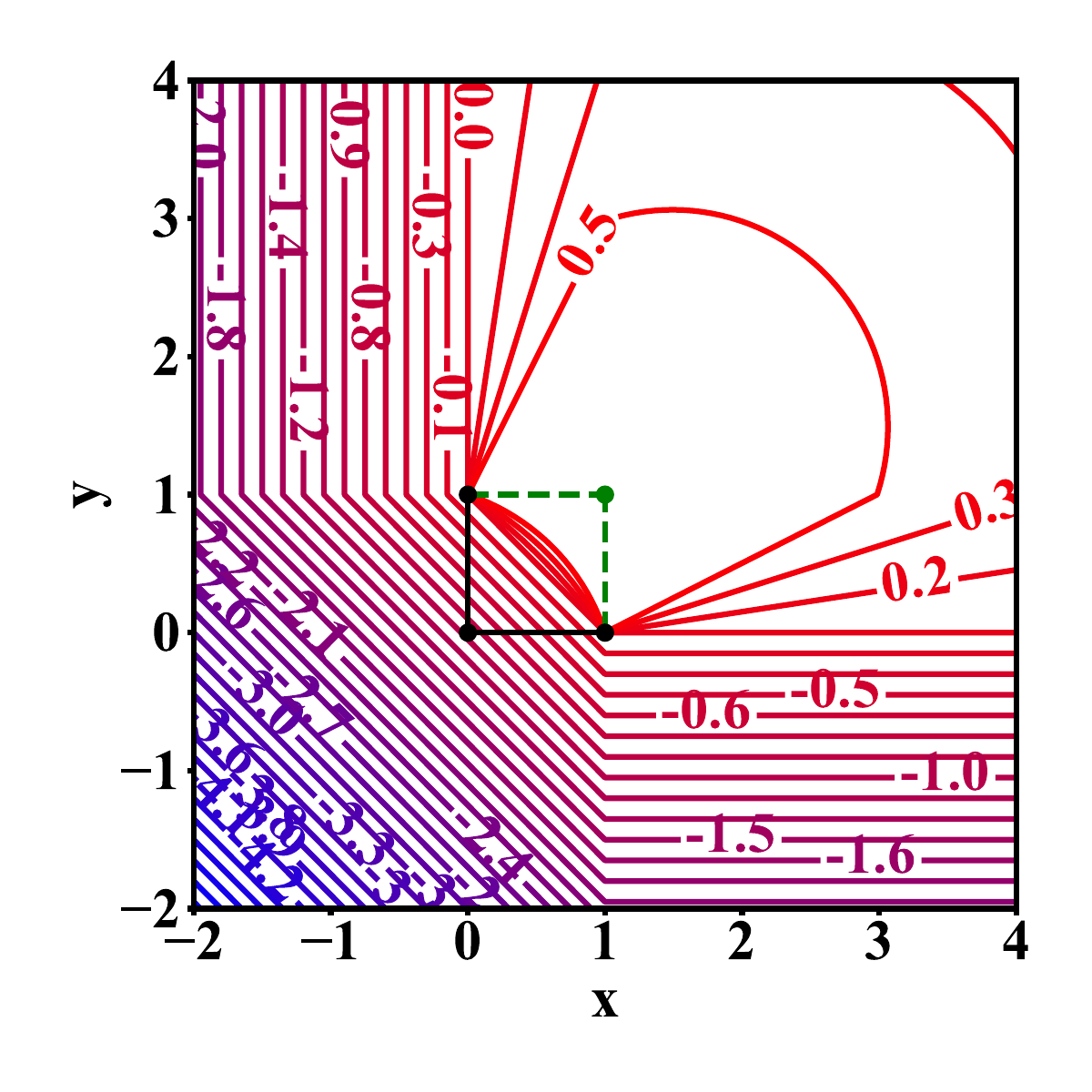}
    \label{fig:2DReHJ}  
    }
  \subfigure[]{
    \includegraphics[width=0.308\textwidth]{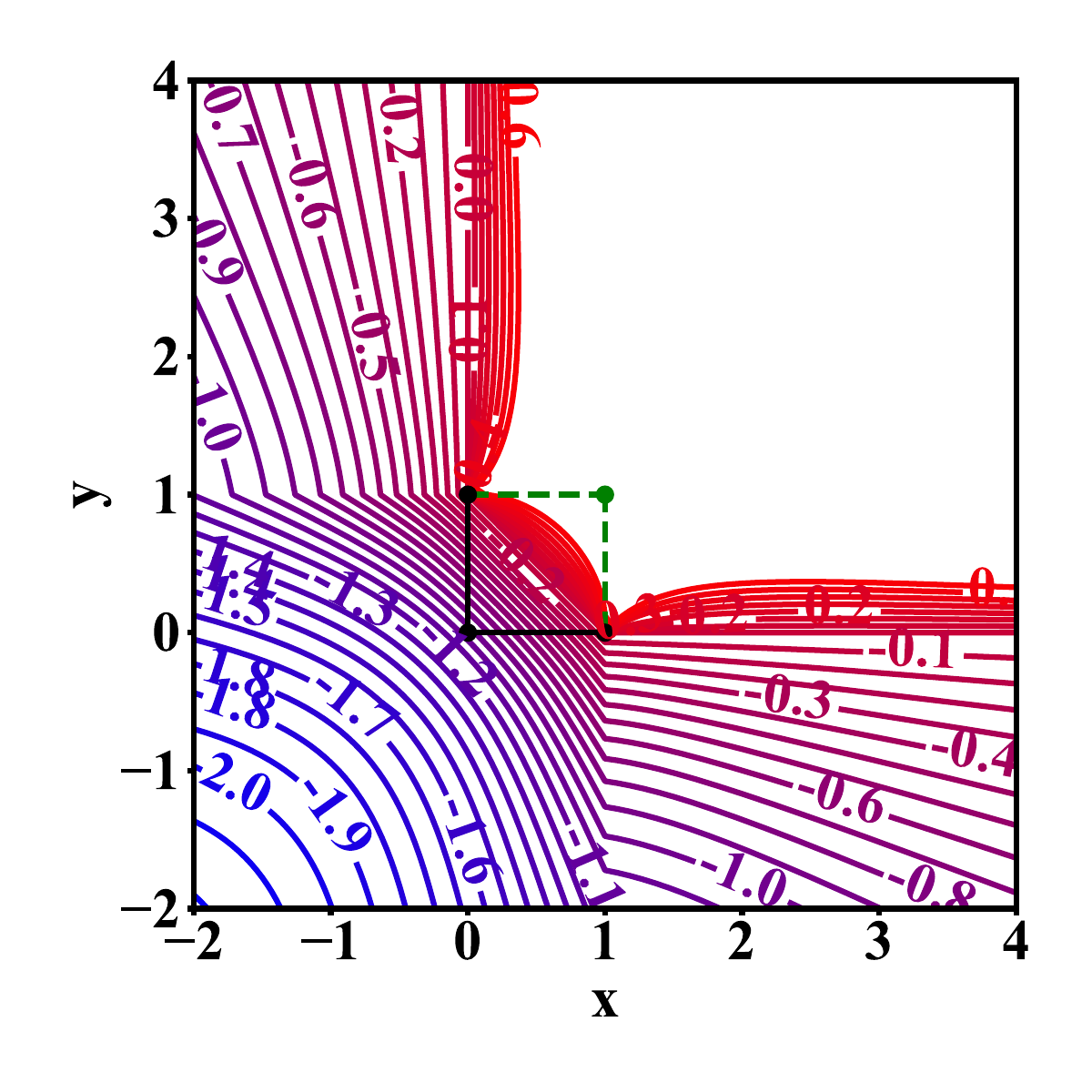}
    \label{fig:2DReHQJ}  
    }
    \vspace{-5mm}
  \caption{ReSJ, ReHJ, and ReHQJ plots of a quadrilateral element with three points (in black) fixed at $(0, 0)$, $(1, 0)$, and $(0, 1)$, and one point (in green) free to move within the plane. $\Theta = 0.6$ is set. (a) The contour plot of function ReSJ. Singular points are encircled with green circles, and local minimum regions are marked with green squares. (b) The contour plot of function ReHJ on the same element. The functional landscape is piecewise linear in the negative Jacobian region. (c) The contour plot of function ReHQJ (the final adopted objective function) on the same element.}
  \label{fig:2DSJJ}  
\end{figure*}

With this setting, it is expected that all hexes with a scaled Jacobian lower than $\Theta$ will be optimized, and the optimization will finish when the objective reaches $N_h\Theta$, where $N_h$ represents the number of hexes in $\mathcal{H}$. However, relying solely on the scaled Jacobian in optimization presents two issues: 1) the scaled Jacobian is non-differentiable at certain points and is non-convex even when only one corner point moves. This problem is illustrated in 2D in Figure \ref{fig:2DSJJ}. In Figure \ref{fig:2DReSJ}, the points encircled by green circles are non-differentiable, and the regions marked by green squares represent local minima. 2) The scaled Jacobian is non-dimensional, and its derivative has an inverse proportional relationship with the hex edge length. Ideally, the derivative should be proportional (i.e., the objective function should be quadratic) to the element size so that the optimization remains invariant to scaling. Otherwise, if the mesh is highly adaptive, elements with a fine scale will have much larger gradients than elements with a coarse scale, as observed in \cite{tong2024hybridoctree_hex}.

To address the first issue, we propose the improved Rectified Hybrid Jacobian (\textit{ReHJ}). As shown in the Jacobian plot in Figure \ref{fig:2DReHJ}, the negative Jacobian region displays a much more convex and everywhere-differentiable landscape, and both the Jacobian and the scaled Jacobian are always either positive or negative. When the Jacobian is non-positive, \textit{ReHJ} is the Jacobian value; when the Jacobian is positive, \textit{ReHJ} is the scaled Jacobian value. We obtain
\begin{equation}
ReHJ(h, \Theta) = \begin{cases}
J(h), & \text{if } J(h) \leq 0 \\
SJ(h), & \text{if } J(h) > 0, SJ(h) \leq \Theta \\
\Theta. & \text{if } SJ(h) > \Theta
\end{cases}
\label{equ:ReHJ}
\end{equation}

To address the second issue, recall that given three edge vectors $e_0, e_1, e_2$ at hex element $h$'s corner/center $x$, we have
\begin{equation}
\begin{aligned}
J(x) &= \det{(e_0, e_1, e_2)},\\
SJ(x) &= \det{\left(\frac{e_0}{\lVert e_0 \rVert_2}, \frac{e_1}{\lVert e_1 \rVert_2}, \frac{e_2}{\lVert e_2 \rVert_2}\right)},
\end{aligned}
\label{equ:JSJOld}
\end{equation}
where $h$'s average edge length is denoted as $\bar{e}$. We scale $J$ and $SJ$ to quadratic measures, namely $QJ$ and $QSJ$, as follows:
\begin{equation}
\begin{aligned}
QJ(x) &= \frac{J(x)}{\Bar{e}},\\
QSJ(x) &= SJ(x)\Bar{e}^2.
\end{aligned}
\label{equ:JSJNew}
\end{equation}
The updated objective function, called Rectified Hybrid Quadratic Jacobian (\textit{ReHQJ}), is written as
\begin{equation}
ReHQJ(h, \theta) = \begin{cases}
QJ(h), & \text{if } J(h) \leq 0 \\
QSJ(h), & \text{if } J(h) > 0, SJ(h) \leq \Theta \\
\Theta, & \text{if } SJ(h) > \Theta
\end{cases}
\label{equ:ReHQJ}
\end{equation}
where it should be noted that since the function of $\bar{e}$ is to normalize the average edge length of $h$ and we do not want to change the landscape of $J$ and $SJ$, $\bar{e}$ is considered as a constant in Equation (\ref{equ:JSJNew}) and does not participate in the gradient calculation. Its plot is shown in Figure \ref{fig:2DReHQJ}.

\subsection{Constraint Setting and Augmented Lagrangian}

\begin{figure}
  \centering
  \includegraphics[width=0.3\textwidth]{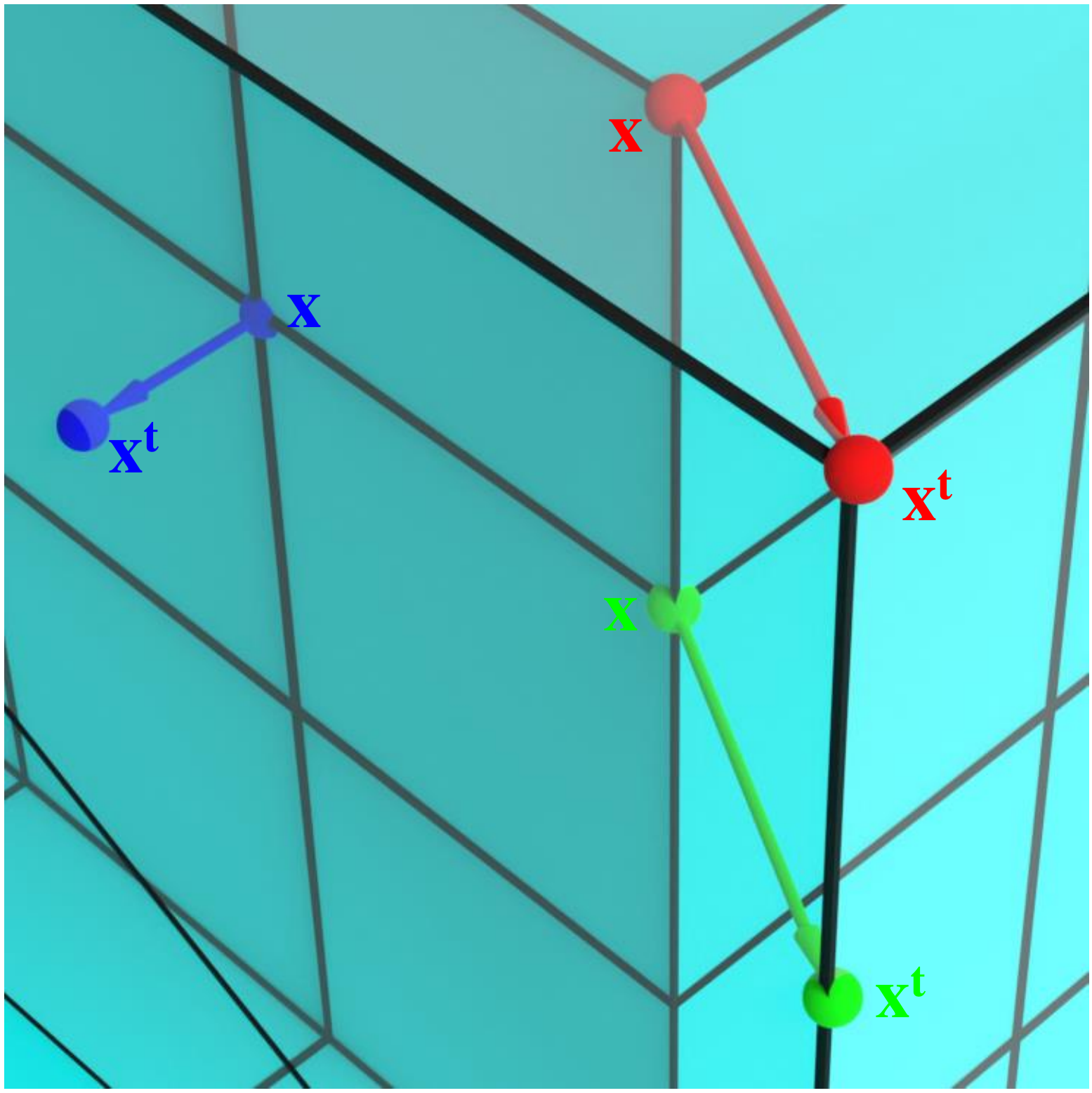}
  \vspace{-2mm}
  \caption{Three types of optimization constraints: (1) the sharp corner point $x$ and its target position $x^t$ in red; (2) the sharp edge point $x$ and its target position $x^t$ in green; and (3) the face point $x$ and its target position $x^t$ in blue.}
  \label{fig:sharpfeature}
\end{figure}

As discussed in Section \ref{sec:1}, the optimization of mesh quality is subject to the constraint that the boundary surface $\mathcal{S}_\mathcal{H}$ must be fitted to the input triangular mesh $\mathcal{T}$. Specifically in Figure \ref{fig:sharpfeature}, for each point $x_i \in \mathcal{S}_\mathcal{H}$, we compute its target point $x_i^t \in \mathcal{T}$ to which $x_i$ should project. The determination of $x_i^t$ depends on the type of features: If $x_i$ is a sharp corner (red), then $x_i^t$ is projected to the corresponding corner point. If $x_i$ is on a sharp edge (green), we compute its projection onto each candidate sharp edge and select the closest projection point as $x_i^t$. If $x_i$ is a face point (blue), we compute its projection onto each triangle and select the closest projection point as $x_i^t$. Once the target points $\{x_i^t\}_{i=1}^{N_s}$ are determined, where $N_s$ denotes the number of vertices in $\mathcal{S}_\mathcal{H}$, these constraints can be formulated as $N_s$ equality conditions. The optimization problem is then formulated as:
\begin{equation}
\begin{split}
\max\sum_{h\in\mathcal{H}}ReHQJ(h, \Theta)\\
\text{subject to }Z_k = Z_k^t,
\label{equ:Opt}
\end{split}
\end{equation}
where $x_i\in\mathcal{S}_\mathcal{H}, Z_k = (x_0, x_1, \cdots, x_{N_s-1})$ in optimization iteration k. $Z_k^t \in \mathcal{T}$ denotes the target points to which $Z_k$ should project. $\Theta = 0$ is set initially, representing the desire for at least an all-positive-Jacobian mesh. Similar to Equation (\ref{equ:naive}), the optimization terminates when $\sum_{h\in\mathcal{H}}ReSJ(h,\Theta) = N_h\Theta$ and all constraints are met. Subsequently, $\Theta$ is incremented by $0.01$, the optimization of $\mathcal{H}$ is repeated with the new configuration, and the previous solution $\mathcal{H}'$ is used as the ``warm start''. This iterative process continues until the optimization problem becomes infeasible.

The constrained optimization problem (\ref{equ:Opt}) can be reformulated as an unconstrained minimization problem through the AL method:
\begin{equation}
\begin{split}
\label{equ:AL}
\min\mathcal{L}(\mathcal{H}, \Theta, \mathcal{T})=\min-\sum_{h\in\mathcal{H}}ReHQJ(h, \Theta)+\\\sum_{Z_k}\left[\lambda_i(Z_k-Z_k^t)+\frac{\rho}{2}(Z_k-Z_k^t)^2\right].
\end{split}
\end{equation}
Following each iteration, the Lagrange multipliers $\lambda_i$ are updated:
\begin{equation}
\label{equ:ALLambda}
\lambda_i = \lambda_i + \rho(Z_k - Z_k^t).
\end{equation}
Each time the minimum scaled Jacobian reaches $\Theta$ (i.e., $\sum_{h\in\mathcal{H}} ReSJ(h,\Theta) = N_h \Theta$), the barrier coefficient $\rho$ is doubled to strengthen the boundary constraints.

\subsection{L-BFGS and Line Search}

An optimizer is essential for determining both the search direction and the step size for $\mathcal{H}$ to minimize $\mathcal{L}$. While a simple approach would be to use Gradient Descent, given the assumption that $\mathcal{S}_\mathcal{H}$ closely approximates $\mathcal{T}$, a quasi-Newton method with quadratic convergence \cite{bertsekas2014constrained} is employed to compute the search direction. In contrast, gradient descent employs the negative gradient $p_k=-\nabla_{x\in\mathcal{H}}\mathcal{L}$ directly to compute the search direction. Consequently, the computational time per iteration for gradient descent is notably shorter compared to quasi-Newton methods. However, due to its first-order convergence, gradient descent needs significantly more iterations than quasi-Newton methods to achieve convergence.

Other quasi-Newton methods, including the Levenberg-Marquardt method \cite{roweis1996levenberg}, face a challenge: the calculation of the Hessian and the inversion of sparse matrices are computationally intensive, leading to a considerable rise in computational time. This level of computational demand is not acceptable for mesh optimization tasks involving millions or more variables. A more advantageous equilibrium between the convergence rate and computational complexity is attained through the L-BFGS algorithm. L-BFGS implicitly approximates the inverse Hessian $H_k^{-1}$ for the $k$-th iteration of the function $p_k=-H_k^{-1}\nabla_{x\in\mathcal{H}}\mathcal{L}$ by utilizing information from the preceding $m$ optimization iterations. For estimating the Hessian, it retains data from prior iterations, specifically $s_k = Z_{k+1} - Z_k$ in optimization iteration $k$ and $y_k = \nabla_{k+1}\mathcal{L} - \nabla_k\mathcal{L}$. Although the determination of the search direction using L-BFGS is approximately twice as slow as gradient descent in experimental settings, it requires about an order of magnitude fewer iterations to reach convergence.

After determining the search direction $p_k$, we need to select a proper step size $a_k$. Consider the Taylor expansion of $\mathcal{L}$ at $Z_k$:
\begin{equation}
\mathcal{L}(Z_k+a_kp_k)=\mathcal{L}(Z_k)+a_kp_k^T\nabla\mathcal{L}(Z_k+ta_kp_k).
\end{equation}
The exact value of $t$ remains unknown, thus we adopt the Amijo line search to numerically determine $t$. We introduce a small parameter $c_1 = 10^{-4} \in (0, 1)$ alongside $\nabla\mathcal{L}(Z_k)$ to relax the constraint on $\mathcal{L}(Z_k + a_k p_k) - \mathcal{L}(Z_k)$. Initially, we set $a_k = 1$ and accept this value if $\mathcal{L}(Z_k + a_k p_k) - \mathcal{L}(Z_k) \leq c_1 a_k p_k^T \nabla\mathcal{L}(Z_k)$. Should this condition not be met, $a_k$ is adjusted through backtracking by multiplying it by $\eta = 0.5$. It is important to note that, theoretically, an arbitrarily small step size could satisfy the aforementioned condition; however, adopting excessively small steps markedly increases the computational time. Consequently, we terminate the backtracking process once $a_k$ drops below $10^{-8}$.

\subsection{Complete Pipeline Pseudo-Code}

The pseudo-code in Algorithm \ref{alg:pipeline} summarizes our proposed pipeline. The process of searching for $x_i^t$ for each $x_i$ involves iterating over all corner points, edges, and faces, which introduces a significant computational overhead. Although confining the search to objects within a predefined searching box centered at $x_i$ typically yields similar $\mathcal{H}'$ outcomes, global traversals are necessary to ensure the highest possible success of the optimization. The constants used in the pipeline are experimentally determined, and variables with subscripts less than zero are disregarded as they represent invalid or initialization states.
\renewcommand{\algorithmicrequire}{\textbf{Input:}}
\renewcommand{\algorithmicensure}{\textbf{Output:}}

\begin{algorithm}[HexOpt]{All-Hexahedral Mesh Quality Improvement}
\label{alg:pipeline}
\begin{algorithmic} [1]
\Require Manifold, watertight triangular mesh $\mathcal{T}$ with annotated sharp features, an all-hex mesh $\mathcal{H}$, minimum scaled Jacobian threshold $\Theta$
\Ensure Warped all-hex mesh $\mathcal{H}'$ with good mesh quality and its boundary $\mathcal{S}_\mathcal{H}$ fitted to $\mathcal{T}$
\State Initialize $N_h \gets \textit{\#elem}\in\mathcal{H}, N_s \gets \textit{\#vert}\in\mathcal{S}_\mathcal{H}, $ iteration number $k \gets 0, $ history length $ m \gets 15, $ Lagrangian multiplier $\lambda \gets 0, $ penalty coefficient $\rho \gets 10^{-8}, $ Armijo constant $c_1 \gets 10^{-4}$
\While{$\exists x_i \in \mathcal{S}_\mathcal{H}, \| x_i - x_i^t \|> 10^{-8}$}
    \State Calculate $x_i^t, \forall x_i \in \mathcal{S}_\mathcal{H}$ \Comment{Update equality constraints}
    \State Call L-BFGS to update variables \Comment{See Algorithm \ref{alg:l_bfgs}}
    \If{$k \% 100 == 0$}
        \State smartLaplacianSmoothing($\mathcal{H}$) \Comment{Smooth the mesh}
    \EndIf
    \State $k \gets k + 1$
\EndWhile
\State \Return $\mathcal{H}'$
\end{algorithmic}
\end{algorithm}

\begin{algorithm}[HexOpt]{L-BFGS Update}
\label{alg:l_bfgs}
\begin{algorithmic} [1]
\Require Current vertices $x_i$, target vertices $x_i^t, \rho, \lambda, N_h, N_s, k, m, c_1$, history vectors $\rho_{i}, s_{i}, y_{i}, i=k-m, k-m+1, \cdots, k-1, i\geq0$
\Ensure Updated variables $Z_{k+1}$
\State Calculate gradient $q \gets \nabla \mathcal{L}_k$ \Comment{See Equation (\ref{equ:AL})}
\For{$i = k - 1, k - 2, \cdots, k - m$} \Comment{First loop}
    \State $\alpha_i \gets \rho_i s_i^T q$
    \State $q \gets q - \alpha_i y_i$
\EndFor
\State $r \gets H_k^0 q = \frac{s_{k-1}^T y_{k-1}}{y_{k-1}^T y_{k-1}} q$
\For{$i = k - m, k - m + 1, \cdots, k - 1$} \Comment{Second loop}
    \State $\beta \gets \rho_i y_i^T r$
    \State $r \gets r + s_i (\alpha_i - \beta)$
\EndFor
\State $Z_k \gets (x_0, x_1, \cdots, x_{N_s-1})$
\State $Z_k^t \gets (x_0^t, x_1^t, \cdots, x_{N_s-1}^t)$
\State $s_k \gets Z_k - Z_{k-1}$
\State $y_k \gets \nabla \mathcal{L}_k - \nabla \mathcal{L}_{k-1}$
\If{$y_k^T s_k == 0$}
    \State $\rho_k \gets 10^8$
\Else
    \State $\rho_k \gets \frac{1}{y_k^T s_k}$
\EndIf
\State $\lambda_i \gets \lambda_i + \rho (Z_k - Z_k^t)$ \Comment{Update Lagrange multiplier}
\If{$\sum_{h\in\mathcal{H}}ReSJ(h,\Theta) = N_h\Theta$}
    \State $\rho \gets 2\rho$ \Comment{Update penalty term}
\EndIf
\State // Armijo Line Search
\State $a_k \gets 1$
\While{$a_k > 10^{-8} \text{ and } \mathcal{L}(Z_k + a_k r) - \mathcal{L}(Z_k) > c_1 a_k r^T \nabla \mathcal{L}(Z_k)$}
    \State $a_k \gets 0.9 a_k$
\EndWhile
\State $(x_0, x_1, \cdots, x_{N_s-1}) \gets Z_k + a_k r$ \Comment{Update variables}
\end{algorithmic}
\end{algorithm}

\section{Results and Discussion}
\label{sec:3}

We evaluate our method on a range of input hexahedral meshes generated using various state-of-the-art algorithms archived in HexaLab \cite{bracci2019hexalab} and from our group's previous work \cite{yu2022hexgen, tong2024hybridoctree_hex} on a PC equipped with a 3.6 GHz Intel i7-12700 CPU and 64GB of memory. These methods include PolyCube-based approaches \cite{guo2020cut, yu2022hexgen} (rkm012\_1, mount2), cross-field-based techniques \cite{li2012all, livesu2020loopycuts} (impeller, mid2Fem), interactive methods \cite{takayama2019dual, zoccheddu2023hexbox} (bunny, CAD4), and octree-based methods \cite{marechal2009advances, tong2024hybridoctree_hex} (anc101, isidore\_horse). Some of these meshes contain inverted elements, while others have all positive Jacobians but deviated surfaces. We intentionally tangle interior vertices of all meshes (boundary vertices remain unmoved) to increase the difficulty of quality improvement.

\begin{table*}
\caption{Hex mesh statistics for models optimized with HexOpt.}
\label{tab:mesh-statistics}
\vspace{2mm}
\centering
\begin{tabular}{cccccccc}
\toprule
Model & \#Vert & \#Elem & OriSJ & PreSJ & PostSJ & PreMaxDist & \makecell{L-BFGS/GD\\Time (s)} \\
\midrule
rkm012\_1&21,312&18,751&[0.47, 1.0]\cite{guo2020cut}&[-1.0, 0.83]&[0.60, 1.0]&0.002072&25/42\\
mount2&7,945&6,208&[0.14, 1.0]\cite{yu2022hexgen}&[-1.0, 0.62]&[0.37, 1.0]&0.005451&10/36\\
impeller&15,248&11,174&[0.18, 1.0]\cite{li2012all}&[-1.0, 0.96]&[0.43, 1.0]&0.0009514&20/48\\
mid2Fem&1,590&908&[-0.15, 1.0]\cite{livesu2020loopycuts}&[-1.0, 0.41]&[0.48, 1.0]&0.0002551&5/10\\
bunny&3,724&2,832&[-0.77, 0.98]\cite{takayama2019dual}&[-1.0, 0.77]&[0.12, 0.98]&0.0&8/18\\
CAD4&3,721&2,704&[0.069, 1.0]\cite{zoccheddu2023hexbox}&[-1.0, 1.0]&[0.12, 1.0]&0.0&9/20\\
anc101&154,675&135,982&[0.017, 1.0]\cite{marechal2009advances}&[-1.0, 0.87]&[0.33, 1.0]&0.003714&69/198\\
isidore\_horse&209,974&182,124&[N/A, 1.0]\cite{tong2024hybridoctree_hex}&[-1.0, 1.0]&[0.54, 1.0]&0.02702&54/171\\
\bottomrule
\end{tabular}    
\begin{minipage}{\textwidth}
\footnotesize{%
  From left to right: model name, number of vertices, number of hex elements, scaled Jacobian range of original models, scaled Jacobian range before optimization, scaled Jacobian range after optimization, maximum relative deviation from $\mathcal{T}$ before optimization (maximum relative deviation from $\mathcal{T}$ after optimization, PostMaxDist, are all 0), L-BFGS/gradient descent running time in seconds.
}  
\end{minipage}
\end{table*}

\begin{figure}
  \centering
  \includegraphics[width=0.47\textwidth]{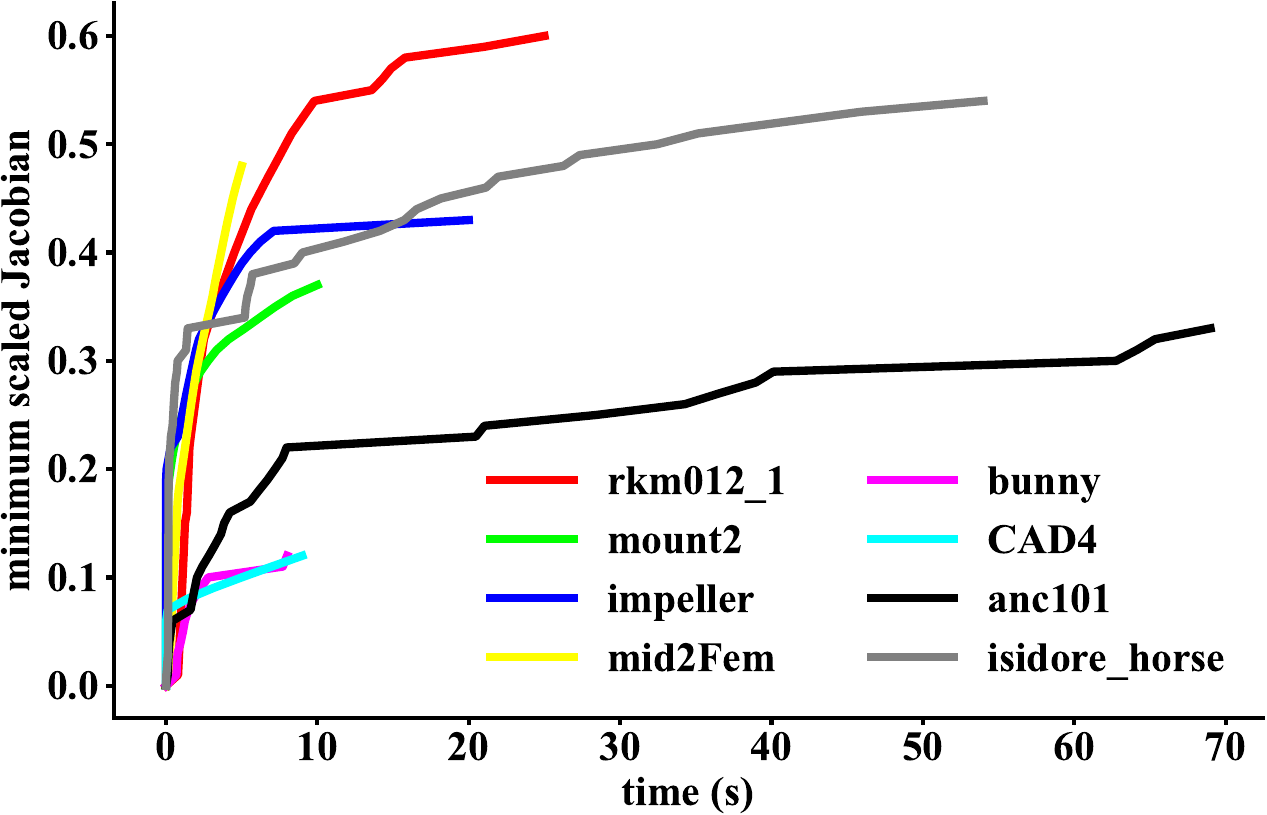}
  \vspace{-2.5mm}
  \caption{Convergence plot on the exhibited eight models.}
  \label{fig:convergence}  
\end{figure}

\begin{figure*}[t!]
  \hspace{-3mm}
  \includegraphics[width=1\textwidth]{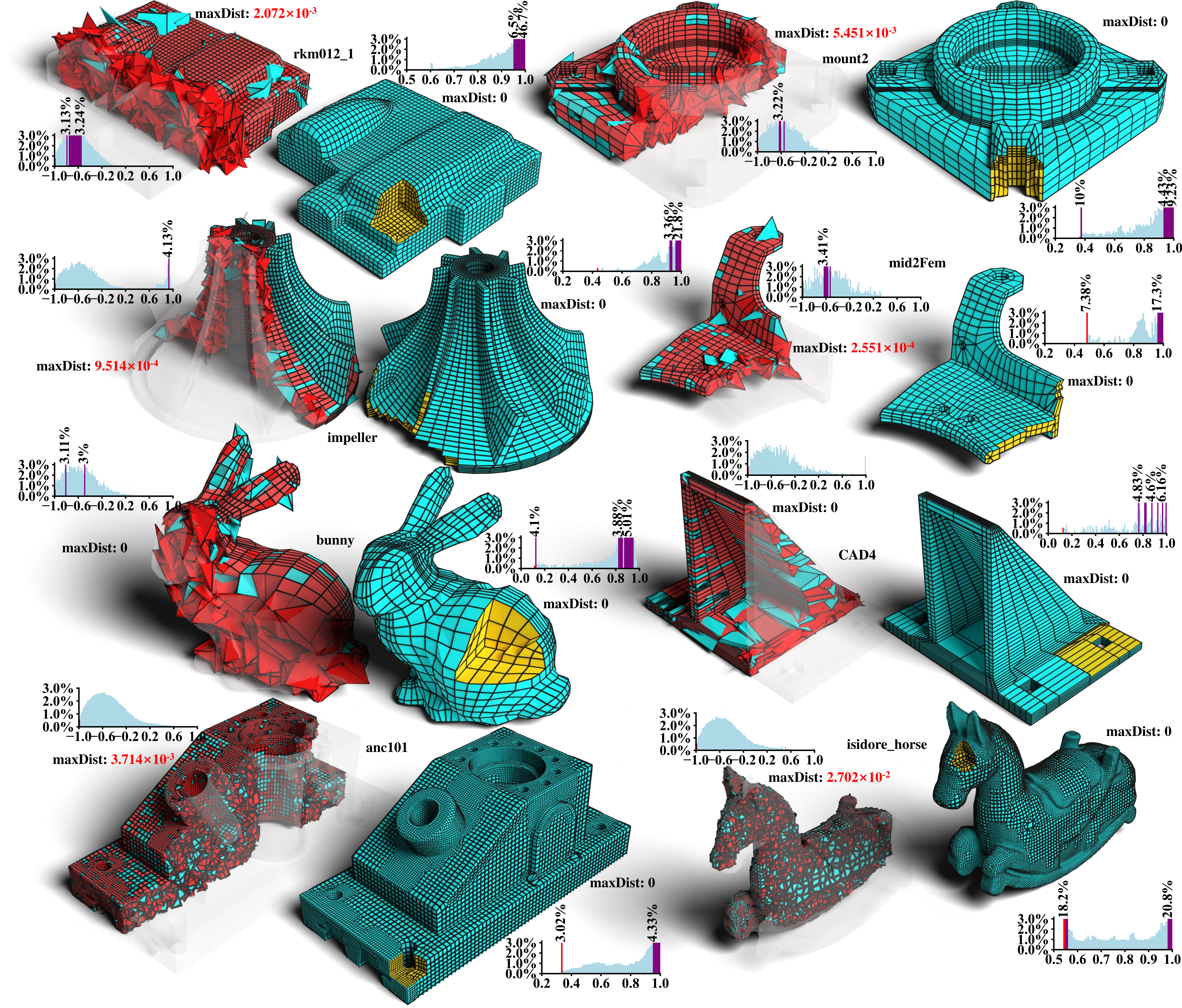}
  \vspace{-3mm}
  \caption{Optimization results for eight models. The target surface is displayed translucently. For each model, the left/right figure shows the mesh before/after optimization. Inverted elements are highlighted in red, and non-inverted elements are shown in blue. The mesh interior is shown in yellow with some elements removed. The maximum relative distance and the scaled Jacobian histogram before and after optimization are provided. In the histograms, purple bars are intersected for better visualization due to their high frequency.}
  \label{fig:result01}
\end{figure*}

As shown in Figure \ref{fig:result01} and Table \ref{tab:mesh-statistics}, HexOpt consistently produces inversion-free hexahedral meshes $\mathcal{H}'$ and significantly improves the worst-scaled Jacobian. We optimize using the perturbed initial mesh, and the minimum scaled Jacobian after optimization always exceeds that of the original initial mesh. The ``maxDist'' value in the figure represents the maximum relative distance. This is calculated by traversing through the vertices in $\mathcal{S}_\mathcal{H}$, finding the vertex with the longest distance to $\mathcal{T}$, and dividing this distance by the diagonal length of $\mathcal{T}$'s bounding box. All the PostMaxDist are 0, representing that the optimized mesh boundary $\mathcal{S}_\mathcal{H}$ remains exactly on the input surface $\mathcal{T}$. Our findings indicate that L-BFGS achieves PostSJ comparable to that of gradient descent while reducing computation time by approximately 50\%.

Additionally, we observe that the method performs particularly well with meshes that have an aspect ratio close to 1 (rkm012\_1, mid2Fem, anc101, and isidore\_horse). A plausible explanation for this phenomenon is that the scaled Jacobian of elements with large aspect ratios is highly sensitive to vertex movements along the shorter edges, making it considerably more challenging to achieve an optimal solution. From this perspective, the most suitable application for HexOpt may be as a post-optimization technique for octree-based meshes. Finally, the convergence plots of the minimum scaled Jacobian for these models are shown in Figure \ref{fig:convergence}. At each save node, \(\mathcal{S}_\mathcal{H}\) is exactly fitted to \(\mathcal{T}\). From the plots, we observe that it typically takes some time to reach convergence. However, the minimum scaled Jacobian is rapidly improved to 60\%–70\% of its final value, taking only around 10\% of the total time. Subsequently, the rate of improvement slows down. Therefore, in practice, one can stop the optimization process once the slope of the convergence curve becomes sufficiently shallow, without necessarily waiting for complete convergence.

We evaluate the necessity of the objective function configuration by substituting Equation (\ref{equ:ReHQJ}) in Equation (\ref{equ:AL}) with Equation (\ref{equ:ReHJ}). The optimization failed to converge for the models anc101 and isidore\_horse due to their surface mesh adaptivity. Furthermore, when Equation (\ref{equ:ReSJ}) was used instead, the optimization failed to converge for all models.

One limitation of HexOpt is its lack of a theoretical lower bound on the minimum scaled Jacobian. We believe that establishing such a bound is a challenging problem, particularly when the shape of \(\mathcal{T}\) is also considered. Another limitation arises from our initial approach of employing path-finding algorithms to detect sharp features, which exhibited unreliability in models with large-aspect-ratio elements proximal to sharp features, such as mount2 and CAD4. This unreliability is caused by boundary quadrilaterals with two adjacent edges lying on a straight path, resulting in a zero scaled Jacobian for the element. Improvement of the Jacobian bound necessitates local padding \cite{marechal2009advances} or pillowing \cite{owen1998survey, qian2012automatic}, which introduces singularity vertices and is unwanted in some applications. Consequently, users need to specify the one-to-one relationship between sharp corners/edges on \(\mathcal{T}\) and \(\mathcal{S}_\mathcal{H}\).

\section{Conclusion and Future Work}
\label{sec:4}

In this paper, we introduce HexOpt, a software package for improving the quality of all-hex meshes. Given a poor-quality or inexact surface-fitting hex mesh and a triangular surface onto which the hex mesh must be projected, HexOpt formulates a constrained optimization problem that includes both mesh quality and geometry fitting terms. The algorithm then employs the AL, L-BFGS and Armijo line search methods to minimize the objective function. This approach is robust, efficient, and fully automated, making it particularly suitable for improving mesh quality for complex 3D models. Across all tested models, selected from our group's previous work and other researchers' archives, our algorithm consistently generates meshes of superior quality. To support further research and advancements in the field, we have made the code and meshing results publicly available at \url{https://github.com/CMU-CBML/HexOpt}.

While HexOpt has demonstrated promising results in rapid, robust, and high-quality all-hex mesh optimization for industrial applications, there is still room for future research. Specifically, establishing theoretical proofs to guarantee optimization performance remains an open challenge. Although some previous methods have proven convergence by optimizing only one vertex at a time to decompose the global optimization problem into local sub-problems \cite{freitag2000local}, or by iteratively moving surface points closer to target points while maintaining positive Jacobians after each step \cite{lin2015quality}, these methods only guarantee that mesh quality improves monotonically. These methods do not provide a lower bound on mesh quality. In practice, they often perform poorly because they impose excessive constraints on the optimization process. Additional mesh generation algorithms that guarantee mesh quality have been developed for quadrilateral meshes \cite{liang2009guaranteed, liang2011hexagon,liang2012matching} and tetrahedral meshes \cite{liang2014octree}. These methods leverage the advantageous topological and geometric properties of 2D Euclidean spaces and 3D simplex. However, adapting these approaches to all-hex meshes presents significant difficulties. Our future efforts will focus on formulating theoretical proofs for the lower bound of mesh quality from a surface mapping perspective and developing more pre-processing functionalities such as local/global pillowing, automatic feature line path finding, and CAD-aware exact geometry fine tuning. In addition, machine learning approaches \cite{tong2023srlafm} could help reduce heuristic steps in mesh generation which is another promising research direction for the future.

\section*{Acknowledgements}
{ 
H. Tong and Y. J. Zhang were supported in part by the NSF grant CMMI-1953323 and a Honda grant.

}

\bibliographystyle{siam}
\bibliography{ltexpprt_references}

\begin{thebibliography}{10}

\bibitem{benzley1995comparison}
{\sc S.~E. Benzley, E.~Perry, K.~Merkley, B.~Clark, and G.~Sjaardama}, {\em A comparison of all hexagonal and all tetrahedral finite element meshes for elastic and elasto-plastic analysis}, in 4th International Meshing Roundtable, 1995, pp.~179--191.

\bibitem{bertsekas2014constrained}
{\sc D.~P. Bertsekas}, {\em Constrained optimization and Lagrange multiplier methods}, Academic Press, 2014.

\bibitem{bracci2019hexalab}
{\sc M.~Bracci, M.~Tarini, N.~Pietroni, M.~Livesu, and P.~Cignoni}, {\em Hexalab. net: An online viewer for hexahedral meshes}, Computer-Aided Design, 110 (2019), pp.~24--36.

\bibitem{canann1998approach}
{\sc S.~A. Canann, J.~R. Tristano, M.~L. Staten, et~al.}, {\em An approach to combined {L}aplacian and optimization-based smoothing for triangular, quadrilateral, and quad-dominant meshes}, International Meshing Rountable, 1 (1998), pp.~479--94.

\bibitem{freitag1997combining}
{\sc L.~A. Freitag}, {\em On combining {L}aplacian and optimization-based mesh smoothing techniques}, Technical Report, Argonne National Lab,  (1997).

\bibitem{freitag2000local}
{\sc L.~A. Freitag and P.~Plassmann}, {\em Local optimization-based simplicial mesh untangling and improvement}, International Journal for Numerical Methods in Engineering, 49 (2000), pp.~109--125.

\bibitem{guo2020cut}
{\sc H.~X. Guo, X.~Liu, D.-M. Yan, and Y.~Liu}, {\em Cut-enhanced polycube-maps for feature-aware all-hex meshing}, ACM Transactions on Graphics, 39 (2020), pp.~106--1.

\bibitem{huang2022untangling}
{\sc Q.~Huang, W.-X. Zhang, Q.~Wang, L.~Liu, and X.-M. Fu}, {\em Untangling all-hex meshes via adaptive boundary optimization}, Graphical Models, 121 (2022), p.~101136.

\bibitem{knupp2001hexahedral}
{\sc P.~M. Knupp}, {\em Hexahedral and tetrahedral mesh untangling}, Engineering with Computers, 17 (2001), pp.~261--268.

\bibitem{knupp2006verdict}
{\sc P.~M. Knupp, C.~Ernst, D.~C. Thompson, C.~Stimpson, and P.~P. Pebay}, {\em The verdict geometric quality library}, tech. rep., Sandia National Laboratories, 2006.

\bibitem{li2012all}
{\sc Y.~Li, Y.~Liu, W.~Xu, W.~Wang, and B.~Guo}, {\em All-hex meshing using singularity-restricted field}, ACM Transactions on Graphics, 31 (2012), pp.~1--11.

\bibitem{liang2009guaranteed}
{\sc X.~Liang, M.~S. Ebeida, and Y.~Zhang}, {\em Guaranteed-quality all-quadrilateral mesh generation with feature preservation}, Computer Methods in Applied Mechanics and Engineering, 199 (2010), pp.~2072--2083.

\bibitem{liang2011hexagon}
{\sc X.~Liang and Y.~Zhang}, {\em Hexagon-based all-quadrilateral mesh generation with guaranteed angle bounds}, Computer Methods in Applied Mechanics and Engineering, 200 (2011), pp.~2005--2020.

\bibitem{liang2012matching}
{\sc X.~Liang and Y.~Zhang}, {\em Matching interior and exterior all-quadrilateral meshes with guaranteed angle bounds}, Engineering with Computers, 28 (2012), pp.~375--389.

\bibitem{liang2014octree}
{\sc X.~Liang and Y.~Zhang}, {\em An octree-based dual contouring method for triangular and tetrahedral mesh generation with guaranteed angle range}, Engineering with Computers, 30 (2014), pp.~211--222.

\bibitem{lin2015quality}
{\sc H.~Lin, S.~Jin, H.~Liao, and Q.~Jian}, {\em Quality guaranteed all-hex mesh generation by a constrained volume iterative fitting algorithm}, Computer-Aided Design, 67 (2015), pp.~107--117.

\bibitem{livesu2020loopycuts}
{\sc M.~Livesu, N.~Pietroni, E.~Puppo, A.~Sheffer, and P.~Cignoni}, {\em Loopycuts: practical feature-preserving block decomposition for strongly hex-dominant meshing.}, ACM Transactions on Graphics, 39 (2020), p.~121.

\bibitem{livesu2015practical}
{\sc M.~Livesu, A.~Sheffer, N.~Vining, and M.~Tarini}, {\em Practical hex-mesh optimization via edge-cone rectification}, ACM Transactions on Graphics, 34 (2015), pp.~1--11.

\bibitem{marechal2009advances}
{\sc L.~Mar{\'e}chal}, {\em Advances in octree-based all-hexahedral mesh generation: Handling sharp features}, in 18th International Meshing Roundtable, 2009, pp.~65--84.

\bibitem{owen1998survey}
{\sc S.~J. Owen}, {\em A survey of unstructured mesh generation technology.}, International Meshing Roundtable, 239 (1998), p.~15.

\bibitem{pietroni2022hex}
{\sc N.~Pietroni, M.~Campen, A.~Sheffer, G.~Cherchi, D.~Bommes, X.~Gao, R.~Scateni, F.~Ledoux, J.~Remacle, and M.~Livesu}, {\em Hex-mesh generation and processing: A survey}, ACM Transactions on Graphics, 42 (2022), pp.~1--44.

\bibitem{qian2012automatic}
{\sc J.~Qian and Y.~Zhang}, {\em Automatic unstructured all-hexahedral mesh generation from {B}-{R}eps for non-manifold {CAD} assemblies}, Engineering with Computers, 28 (2012), pp.~345--359.

\bibitem{qian2010quality}
{\sc J.~Qian, Y.~Zhang, W.~Wang, A.~C. Lewis, M.~S. Qidwai, and A.~B. Geltmacher}, {\em Quality improvement of non-manifold hexahedral meshes for critical feature determination of microstructure materials}, International Journal for Numerical Methods in Engineering, 82 (2010), pp.~1406--1423.

\bibitem{roweis1996levenberg}
{\sc S.~Roweis}, {\em Levenberg-{M}arquardt optimization}, Notes, University of Toronto, 52 (1996).

\bibitem{ruiz2015simultaneous}
{\sc E.~Ruiz-Giron{\'e}s, X.~Roca, J.~Sarrate, R.~Montenegro, and J.~M. Escobar}, {\em Simultaneous untangling and smoothing of quadrilateral and hexahedral meshes using an object-oriented framework}, Advances in Engineering Software, 80 (2015), pp.~12--24.

\bibitem{schneiders2000algorithms}
{\sc R.~Schneiders}, {\em Algorithms for quadrilateral and hexahedral mesh generation}, Proceedings of the VKI Lecture Series on Computational Fluid Dynamic,  (2000).

\bibitem{shepherd2008hexahedral}
{\sc J.~F. Shepherd and C.~R. Johnson}, {\em Hexahedral mesh generation constraints}, Engineering with Computers, 24 (2008), pp.~195--213.

\bibitem{shepherd2006quality}
{\sc J.~F. Shepherd, C.~J. Tuttle, C.~Silva, and Y.~Zhang}, {\em Quality improvement and feature capture in hexahedral meshes}, Technical Report UUSCI-2006-029, The University of Utah,  (2006).

\bibitem{takayama2019dual}
{\sc K.~Takayama}, {\em Dual sheet meshing: An interactive approach to robust hexahedralization}, Computer graphics forum, 38 (2019), pp.~37--48.

\bibitem{tong2024hybridoctree_hex}
{\sc H.~Tong, E.~Halilaj, and Y.~J. Zhang}, {\em Hybrid{O}ctree\_{H}ex: Hybrid octree-based adaptive all-hexahedral mesh generation with {J}acobian control}, Journal of Computational Science, 78 (2024), p.~102278.

\bibitem{tong2023srlafm}
{\sc H.~Tong, K.~Qian, E.~Halilaj, and Y.~J. Zhang}, {\em S{RL}-{A}ssisted {AFM}: Generating planar unstructured quadrilateral meshes with supervised and reinforcement learning-assisted advancing front method}, Journal of Computational Science, 72 (2023), p.~102109.

\bibitem{wang2021structure}
{\sc R.~Wang, Z.~Zheng, W.~Yu, Y.~Shao, and S.~Gao}, {\em Structure-aware geometric optimization of hexahedral mesh}, Computer-Aided Design, 138 (2021), p.~103050.

\bibitem{xu2018hexahedral}
{\sc K.~Xu, X.~Gao, and G.~Chen}, {\em Hexahedral mesh quality improvement via edge-angle optimization}, Computers \& Graphics, 70 (2018), pp.~17--27.

\bibitem{yu2022hexgen}
{\sc Y.~Yu, X.~Wei, A.~Li, J.~G. Liu, J.~He, and Y.~J. Zhang}, {\em Hex{G}en and {H}ex2{S}pline: Polycube-based hexahedral mesh generation and spline modeling for isogeometric analysis applications in {LS-DYNA}}, in Geometric Challenges in Isogeometric Analysis, Springer, 2022, pp.~333--363.

\bibitem{zhang2013challenges}
{\sc Y.~Zhang}, {\em Challenges and advances in image-based geometric modeling and mesh generation}, Image-Based Geometric Modeling and Mesh Generation,  (2013), pp.~1--10.

\bibitem{zhang2016geometric}
{\sc Y.~Zhang}, {\em {Geometric Modeling and Mesh Generation from Scanned Images}}, CRC Press, Taylor \& Francis Group, 2016.

\bibitem{zhang2006adaptive}
{\sc Y.~Zhang and C.~Bajaj}, {\em Adaptive and quality quadrilateral/hexahedral meshing from volumetric data}, Computer Methods in Applied Mechanics and Engineering, 195 (2006), pp.~942--960.

\bibitem{zhang20053d}
{\sc Y.~Zhang, C.~Bajaj, and B.-S. Sohn}, {\em 3{D} finite element meshing from imaging data}, Computer Methods in Applied Mechanics and Engineering, 194 (2005), pp.~5083--5106.

\bibitem{zhang2009surface}
{\sc Y.~Zhang, C.~Bajaj, and G.~Xu}, {\em Surface smoothing and quality improvement of quadrilateral/hexahedral meshes with geometric flow}, Communications in Numerical Methods in Engineering, 25 (2009), pp.~1--18.

\bibitem{zhang2010automatic}
{\sc Y.~Zhang, T.~J. Hughes, and C.~L. Bajaj}, {\em An automatic 3{D} mesh generation method for domains with multiple materials}, Computer Methods in Applied Mechanics and Engineering, 199 (2010), pp.~405--415.

\bibitem{zoccheddu2023hexbox}
{\sc F.~Zoccheddu, E.~Gobbetti, M.~Livesu, N.~Pietroni, and G.~Cherchi}, {\em Hexbox: interactive box modeling of hexahedral meshes}, Computer Graphics Forum, 42 (2023), p.~e14899.

\end{thebibliography}

\end{document}